\documentclass{osa-article}

\journal{oe}

\articletype{Research Article}

\begin{document}

\title{Degenerate optical resonator for the enhancement of large laser beams}

\author{Nicolas Mielec\authormark{1,$\dagger$}, Ranjita Sapam\authormark{1,$\dagger$}, Constance Poulain\authormark{1,2}, Arnaud Landragin\authormark{1}, Andrea Bertoldi\authormark{3}, Philippe Bouyer\authormark{3}, Benjamin Canuel\authormark{3} and Remi Geiger\authormark{1,*} }

\address{
\authormark{1}LNE-SYRTE, Observatoire de Paris-Universit\'e PSL, CNRS, Sorbonne Universit\'e,  61 avenue de l'Observatoire, 75014 Paris, France.\\
\authormark{2}Ecole polytechnique, Institut Polytechnique de Paris, Route de Saclay, 91120 Palaiseau, France.\\
\authormark{3}LP2N, Laboratoire Photonique, Num{\'e}rique et Nanosciences, Universit{\'e} Bordeaux--IOGS--CNRS: UMR 5298, rue F. Mitterrand, 33400 Talence, France.\\
}
\authormark{$\dagger$} These authors contributed equally.

\authormark{*}Corresponding author: \email{remi.geiger@obspm.fr}



\begin{abstract}
Enhancement cavities where a beam of large  size (several millimeters) can resonate have several applications, in particular in atomic physics. However, reaching large beam waists in a compact geometry (less than a meter long) typically brings the resonator close to the degeneracy limit. Here we experimentally study a degenerate optical cavity, 44-cm long and consisting of two flat mirrors placed in the focal planes of a lens, in a regime of intermediate finesse ($\sim 150$). We study the impact of the longitudinal misalignement on the optical gain, for different input beam waists up to 5.6~mm, and find data consistent with the prediction of a model based on ABCD propagation of Gaussian beams. 
We reach an optical gain of 26 for a waist of 1.4~mm, which can have an impact on several applications, in particular  atom interferometry.
We numerically investigate the optical gain reduction for large beam waists using the angular spectrum method to consider the effects of optical aberrations, which play an important role in such a degenerate cavity. Our calculations quantitatively reproduce the experimental data and will provide a key tool for designing enhancement cavities close to the degeneracy limit. 
As an illustration, we discuss the application of this resonator geometry to the enhancement of laser beams with top-hat intensity profiles.
\end{abstract}


\section{Introduction}
\label{introduction}
Optical cavities represent one of the most employed optical systems, in use by various scientific communities. 
Beyond their core function in laser resonators, optical cavities have numerous applications, such as in laser interferometry based gravitational wave detectors \cite{Mours2006},  frequency metrology \cite{Jiang2011}, cavity quantum electrodynamics \cite{Cox2018}, ultrafast science \cite{Carstens2013}, and nuclear fusion \cite{Fiorucci2018}.
One essential aspect in several applications is the ability of optical cavities to enhance the power of the light when the resonance condition is maintained. 
In the most common case, optical resonators are formed by a set of mirrors, whose curvature defines the geometry and spectral properties of the resonating mode, whereas the optical enhancement factor  is  essentially determined by the losses of the optics.
In some applications, it is desirable to obtain  resonating beams with large diameters (several mm or cm), which translates in increased resonator length  and radii of curvature of the mirrors (up to km). 
Examples of such optical resonators able to sustain large beams range from gravitational wave detectors to ultrafast science, fusion reactors and antimatter experiments \cite{Ahmadi2018}. 

Enhancement cavities have been proposed for cold-atom interferometry experiments, as a way to relax the requirements on the laser power required to operate atom interferometers \cite{Hamilton2015,Riou2017,Dovale-Alvarez2017,Canuel2018}. This application typically requires two counter-propagating beams with the same spatial profile and with diameters of several mm, in order to efficiently interrogate clouds of atoms at $\mu$K temperatures   which freely propagate in vacuum for hundreds of ms \cite{Geiger2020}.
Reaching such large beam sizes in compact setups ($<1$~m) introduces severe constraints on the resonator design. 
Large mode enhancement cavities have been studied in various configurations. For example, optical power enhancement factors of 2000 were demonstrated for a beam radius of 5~mm in Ref.~\cite{Carstens2013} which used a four-mirror cavity. Ref.~\cite{Fiorucci2018} studied a  three-mirror resonator in a telescope-based configuration achieving beam waists of about 2~mm. 
In the context of atom interferometry, the  need to accommodate the enhancement  cavity in the vacuum system enclosing the atom interferometer leads to considering linear resonators as the easiest-to-implement configuration. For that purpose, Ref.~\cite{Riou2017,patentCavity} proposed to employ a linear, degenerate resonator consisting of two mirrors located at the focal planes of a lens to achieve large beam waists in a compact geometry. This cavity configuration shares similarity with the resonator used in Refs.~\cite{Nixon2013,Chriki2015,Chriki2018} where the authors aim at producing a laser of low coherence by allowing many transverse modes to lase simultaneously.


In this paper, we report on our experimental and numerical study of the degenerate optical resonator proposed in Ref.~\cite{Riou2017} in a compact geometry of 44~cm. 
In such a resonator, an arbitrary input field distribution should reproduce after two round-trips, so that  a large beam  can in principle resonate when a laser beam of large waist in injected in the cavity.
However, due to its degeneracy, achieving proper resonance of a Gaussian beam critically depends on the alignment of the resonator and of the optical quality of the optics. Such requirements motivates the  experimental and numerical investigations that we present for beam diameters up to 11~mm, which are of interest for various applications, including cold-atom interferometry.

The article is organized as follows: we first present the experimental implementation of the resonator and detail the alignment procedure. We then study the influence of longitudinal misalignments on the cavity resonance and its optical enhancement factor, with the support of a model based on ABCD matrix propagation.  We then show experimentally  how   wave front distortions introduced by the optical elements influence the performance of the resonator (optical gain and shape of the beam) when targeting larger resonating beam sizes, and compare the data with a refined model of our cavity, based on the method of the angular spectrum propagation of fields. We finally conclude by summarizing our findings and presenting the perspectives of this study.


\section{Experimental setup and alignment procedure}

\subsection{Optical resonator setup}
\label{subsec:description}

\begin{figure}[h!]
\centering
\includegraphics[width=\linewidth]{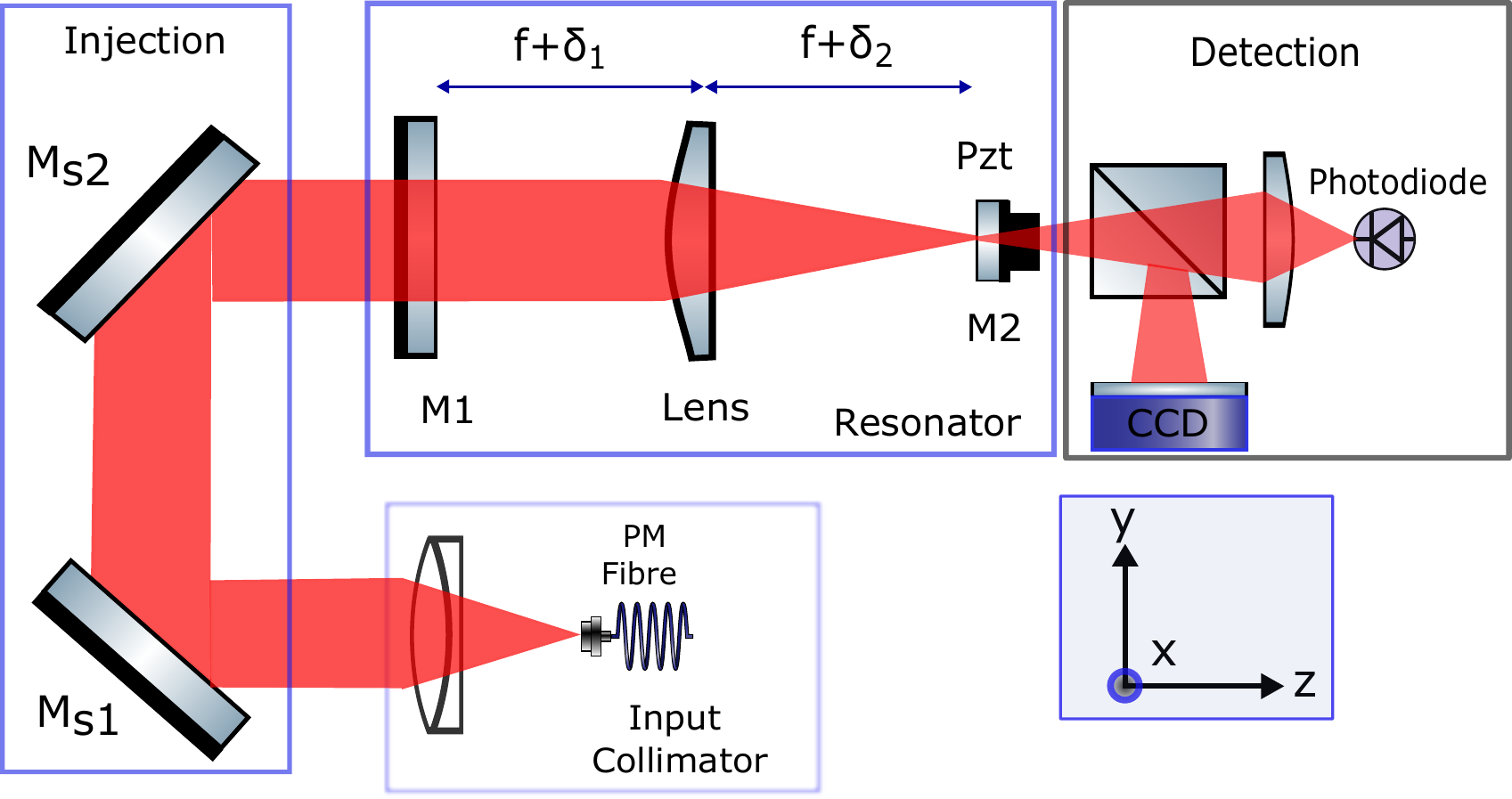}
\caption{Schematic of the degenerate optical resonator setup (simplified and not to scale) consisting of two flat mirrors M1 (input) and M2 (output) at the focal planes of  a plano-convex lens of focal distance $f = 228$~mm. $\delta_ {1,2}$ are the longitudinal misalignment of mirrors M1 and M2 from the focal planes. M$_{\textrm{s1}}$ and M$_{\textrm{s2}}$ are the two input beam steering mirrors, PZT is a cylindrical  piezo electric actuator tube attached to the back of M1.}
\label{fig:schem}
\end{figure}

A simplified view of our experimental set-up  is illustrated in Fig.~\ref{fig:schem}. The resonator  is made of two plane mirrors of diameter 50.8~mm (M1) and 12.7~mm (M2), placed approximately in the focal planes of a plano-convex lens of focal distance $f$. The two mirrors are spaced by $L\simeq 2 f$.
 The mirrors (from Laseroptik) have a  specified surface flatness of $\lambda/10$ peak-valley. They are coated for   high reflectivity  on one surface (IBS coating B-14353), and have an anti-reflection coating on the polished backside. The mirrors have a wedge of  $30'$ and a thickness of 6.5~mm.
The lens is a catalogue product (model  PLCX-50.8-103.0-UV-700-900 from CVI) with 50.8~mm diameter and focal distance $f=228$~mm at the wavelength $\lambda=852$~nm.  
The specified centration error is $66 \ \mu$m.
The optical components are mounted on a massive aluminum plate using precision kinematic  mounts and differential translation stage of $\mu$m resolution for fine tuning.
A collimated Gaussian beam of variable waist, $ w_{in}$  is sent into the cavity from  M1 and is focused on M2 by the  lens with a waist, $ w_{out}= \lambda f/\pi  w_{in}$. In this study, the input waist will be varied between three different values, $ w_{in}=1.4, \ 2.6, \ 5.6$~mm, which are, in particular, relevant to cold-atom interferometry applications \cite{Riou2017}.

The laser beam is generated from an  Extended Cavity Diode Laser (ECDL, linewidth $\sim 100$ kHz)  locked to the D2 line of Cesium by saturated absorption spectroscopy \cite{Baillard2006}. The different values of the input waist  are obtained using three different beam expanders with the same polarization maintaining fiber. The transmitted intensity from the cavity is measured by scanning the PZT voltage attached to  M2 with a linear  ramp. 
The  beam transmitted from the cavity is split in  two beams, in order to record  the intensity variations with a photodiode and the transverse profile (in $(xy)$ plane) by means of a CCD camera.

Denoting as $E_{in}$  the incident field in the cavity,   $E_{cir}$ as the circulating field inside the cavity,  the optical gain is given by $G =|\frac{E_{cir}}{E_{in}}|^2$.  The theoretical expression of $G$ and finesse ,$\mathcal{F}$ evaluated at a plane between the lens and M2  are given by 
\begin{subequations}\label{eqn:GF}
\begin{equation}
G=\frac{r_2(1-r_1^2)(1-r_L^2)}{[1-r_1r_2(1-r_L^2)]^2}
\end{equation}
\begin{equation}
\mathcal{F}=\frac{\pi\sqrt{r_1r_2(1-r_L^2)}}{1-r_1r_2(1-r_L^2)}
\end{equation}
\end{subequations}
where $r_1,r_2 $ and $r_L$ are the amplitude reflection coefficients of M1, M2 and lens, respectively. 
We measured  $r_1=r_2=0.994 \pm 0.005$ for the mirrors;
$r_L$ was affected by a larger uncertainty, and since its value has a strong impact on $\mathcal{F}$ and $G$, we adjusted it using the measured values of  $\mathcal{F}=156\pm 5$ and $G=26\pm 1$  described in section \ref{subsec:gain_finesse} below, corresponding to $r_L$ ranging from 0.086 to 0.095.

\subsection{Alignment Procedure}
\label{subsec:alignement}

\begin{figure}[h!]
\centering
\includegraphics[width=0.7\linewidth]{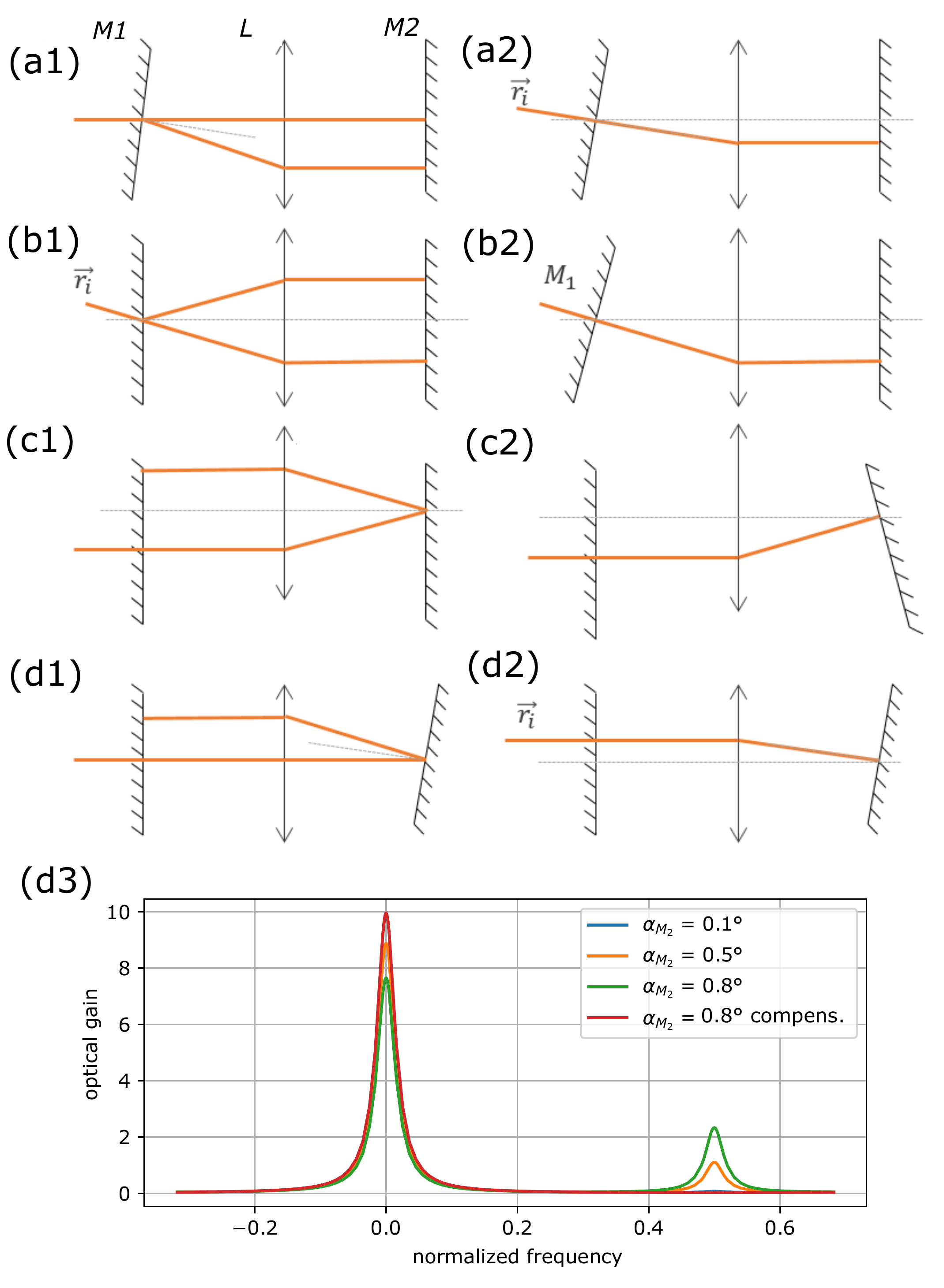}
\caption{Illustration of the effect of different types of transverse misalignment via ray tracing of the center of the optical beam through the cavity. 
The left panels show different types of misalignment, either where an optical element is misaligned (M1 tilted in (a1) or M2 tilted in (d1)), or where the input beam is introduced with a tilt with respect to the optical axis (b1), or with a transverse displacement (c1). The right panels (a2,b2,c2,d2) show how the first misalignment can be compensated by the introduction of a second misalignment. For example, (d2) shows that translating the input beam can compensate the tilt of M2.
The last panel (d3) illustrates how the presence of a misalignment induces a drop (increase) in the amplitude of the even (odd) mode. We illustrate  here  the case where M2 is tilted by a variable angle $\alpha_{M2}$ (situation (d1)). The red line ("compensated") shows how full resonance of the even mode can be recovered by translating the input beam in the transverse direction (situation (d2)). The calculation is performed here in one dimension using the angular spectrum method (see text for details) for a Gaussian beam of 5~mm waist and a resonator with a maximum gain of 10.}
\label{fig:alignement}
\end{figure}

Due to the degeneracy of our resonator, the spectral and spatial features  of the beam resonating in the cavity will critically depend on the relative alignment of  the optical elements  and of the geometrical injection parameters. The theoretical response  of the resonator  to various types of misalignment drove our alignment procedure. 
Through this procedure, we aim at reaching the maximum optical gain, which is theoretically expected when the input beam is aligned with the  optical axis corresponding to the situation where the normal to the surface of the lens and that of M1 and of M2 coincide. 
Precentering of the  optical components is obtained within $\sim 500 \ \mu$m precision by the mechanical design. 
The lens is mounted in a combination mount of precision kinematic tip-tilt (Thorlabs KS2D) with a  resolution of $25 \mu$m/rev, and a $x-y$ translation stage with $250 \ \mu$m/rev (Thorlabs LM2XY). The mirrors are also mounted on precision kinematic tip-tilts.

Initially, M2 is aligned by  auto-collimation with the input beam passing through an iris (not shown in Fig.~\ref{fig:schem}), which constrains input tilts to about 0.5~mrad. 
Then, the  Fabry-Perot cavity made   M1 and M2 is aligned  by using an input beam  of  1~mm diameter and by overlapping on M2   the spots corresponding to the successive reflections on M1. Using a camera, we can overlap with $\simeq 200 \ \mu$m resolution the four spots  associated to four round trips (which corresponds to a total propagation distance of about $8 f\simeq 1.8$~m), by tip-tilting M1, without touching M2. M1 and M2 are  therefore  aligned relative to each other with a resolution of about  $100 \ \mu$rad. 
 Finally, the lens is introduced approximately at its focal distance from M1 and M2. Looking with the camera at the focused spots on M2, we optimize the overlapping of the successive reflections by translating the lens in the $(xy)$ plane, and by tip-tilting it. This yields a first resonance signal observed on the photo detector. 
 The transmitted signal is improved by longitudinally displacing  M2, with variations of $\delta_2$ of typically a few $100 \ \mu$m. 
 
At that stage, the resonance spectrum essentially features two peaks corresponding to the odd and even Laguerre modes of the cylindrically symmetric resonator.
As illustrated in Fig.~\ref{fig:alignement}, transverse misalignment of the optics or of the input beam favor one mode with respect to the other, and can compensate each other to favor only one.
 Via tip-tilt and $xy$ translations on the lens, and tip-tilt on M1, we minimize the odd mode within the resolution of the photo-detector setup: in practice, no influence is observed for tip and tilts below $5 \ \mu$rad (2 degrees rotation of the differential KS2D adjuster screw) and $xy$ translations of less than $30\ \mu$m.

In the last, fine-tuning step of the alignment, we use our knowledge of the behavior of the degenerate resonator \cite{Riou2017}: when M2 is not at the focal distance, the resonance peak becomes asymmetric, as the resonator is not yet fully degenerate. Targeting a symmetric resonance yields a degenerate resonator  corresponding to the $f-f$ configuration, within a resolution of typically $5\ \mu$m on $\delta_2$. 

In this work, we study the behavior of the cavity for different values of the input beam waist, by changing the lens situated after the  collimator (see Fig.~\ref{fig:schem}). Changing this lens affects the trajectory of the input beam onto the cavity, which we compensate by adjusting the beam steering mirrors $M_{s1}$ and $M_{s2}$ to maximize the optical gain. During the last fine optimization stage, we slightly adjust the optics of the resonator to reach the maximum gain.

\section{Results}

\subsection{Influence of longitudinal misalignment}

\subsubsection{Spectral behavior}

\paragraph{Small beam size ($w_{in}=$1.4~mm).}

\begin{figure}[!h]
\centering
\includegraphics[width=\linewidth]{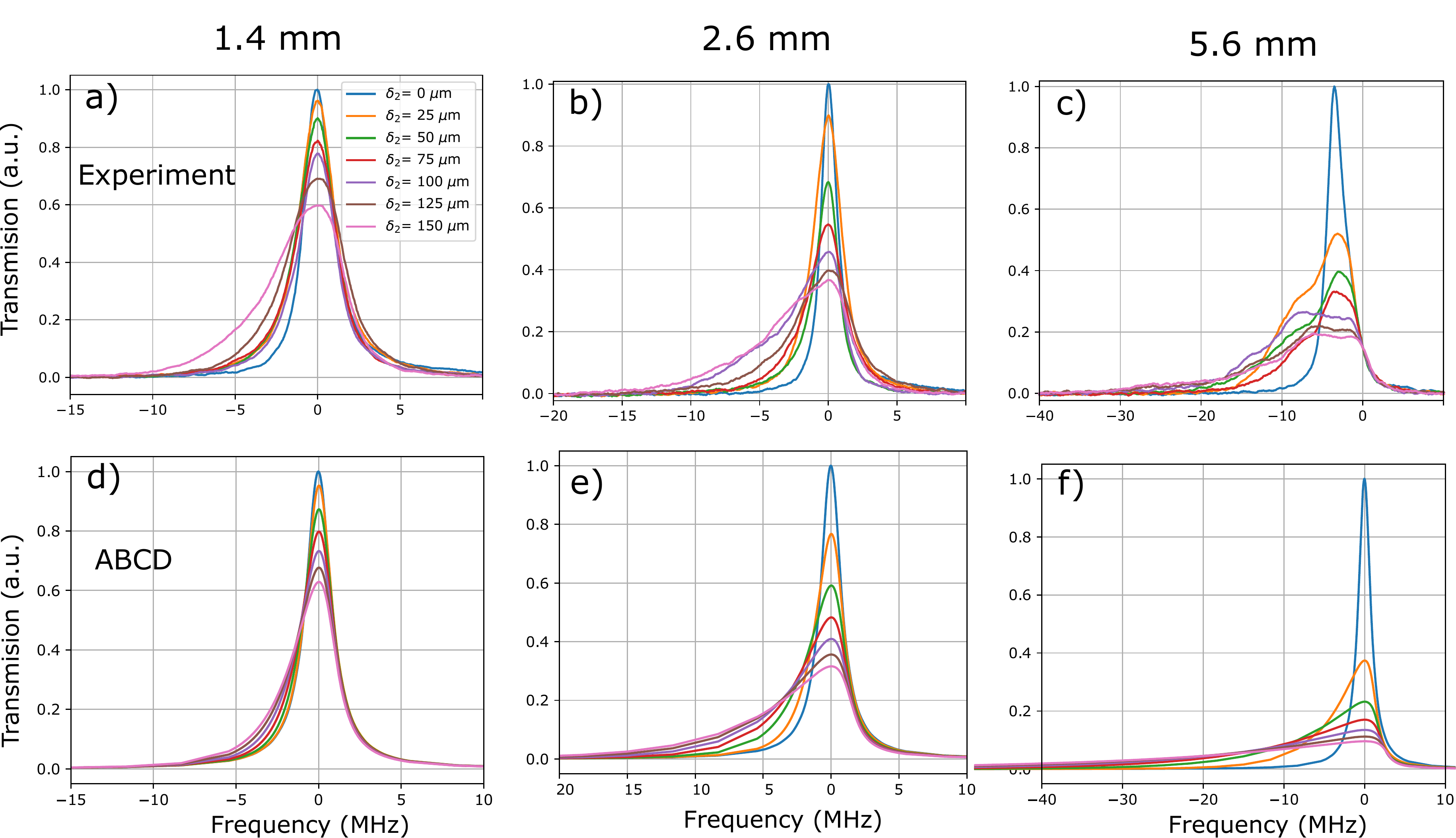}
\caption{Transmission spectra for different values of the longitudinal misalignement $\delta_2$. (a)-(c) are experimental  spectra for input beam waists $w_{in}=1.4, \, 2.6, \ 5.6$~mm, respectively. (d)-(f) are the corresponding spectra calculated  with the ABCD transfer matrix method and assuming aberration-free optics. 
}
\label{fig:resonances}
\end{figure}

To analyze  the behavior of our resonator, we measured its  transmission spectrum for different values of its length around the $f-f$ configuration. The resonances were measured by scanning the voltage of the PZT actuator holding $M_2$ (resulting in length variations of the order of half the wavelength, i.e. $400$~nm). Different resonance curves were recorded  for different values of $\delta_2$.
Fig.~\ref{fig:resonances}(a) shows the  variations in the transmission spectrum  for an input beam of $w_{in}$=1.4 mm, for different values of $\delta_2>0$. 
When increasing $\delta_2$, the spectrum becomes more and more asymmetric and features  a long tail on one side.  

Fig.~\ref{fig:resonances} d) is the corresponding theoretically calculated spectra  using the ABCD transfer matrix method presented in Ref.~\cite{Riou2017}. 
The calculation reproduces well  the broadening and the asymmetric feature of the  resonance, which can  be interpreted with the standard modal theory of optical resonators \cite{Siegman1986} as follows:
in the case of a stable resonator, the  frequencies of the eigenmodes indexed  by $\{q,m,m\}$ ($q$ for  longitudinal, $(m,n)$ for  transverse)   are given by
\begin{equation}
\label{eq:dnu}
\nu_{qmn}=\frac{c}{2L}\left(q+(m+n+1)\frac{\phi_G}{\pi}\right),
\end{equation}
where $\phi_G$ is the Gouy phase.
In the case of our resonator, the perfect $f-f$ configuration ($\delta_2 = 0$) implies $\phi_G = \pi/2$ \cite{Riou2017}. 
Any combination of even modes ($m+n$ even)  resonates at one set of frequencies, and any combination of odd modes ($m+n$ odd) at another, shifted by $c/4L$.

When the resonator is  off from  the $f-f$ configuration ($\delta_2 \neq 0$),  the degeneracy between the transverse and longitudinal modes is lifted, and a fundamental mode ($m=n=0$) is well defined, corresponding to $\phi_G \neq \pi/2$.
 Injecting the cavity with a beam of different waist than that of the resonator  mode results in a decomposition   of the input beam over several transverse modes, which resonate at different frequencies according to Eq.~\eqref{eq:dnu}.
When  $\delta_2\ll f$ is small enough for the  different resonance curves to partially overlap within the resonance linewidth (given by the resonator finesse), the individual resonances corresponding to each transverse mode  cannot be resolved, and the spectrum effectively looks continuous.
The asymmetry of the resonance  is linked to the stability diagram of our cavity which is  similar to that of  a confocal resonator: for a given sign of $\delta_2$ (e.g. $\delta_2<0$, corresponding to $d_2<f$), modifying the length of the cavity yields a stable configuration only for one direction ($d_1<f$), while in the other direction ($d_1>f$) the cavity is unstable.

\paragraph{Larger beam sizes ($w_{in}=$2.6 mm \& 5.6 mm).}
The resonance spectra for $w_{in}$=2.6~mm and 5.6~mm   are shown  in Fig.~\ref{fig:resonances} (b) and (c), respectively.
 We observe that the spectrum has a higher degree of asymmetry and a lower optical gain than for the case of $w_{in}=1.4$~mm, for equivalent values of displacement $\delta_2$. This behavior is qualitatively reproduced by the ABCD matrix calculation (panels (e) and (f)), and can be explained by the higher number of transverse modes over which the input beam  projects when the input waist increases, for a given value of $\delta_2$.
 
For the largest input beam size,  we observe the appearance of  additional structures in the resonance profile (dips and bumps on the elongated tail), which are not captured by the ABCD matrix calculation. Extending the ABCD matrix propagation to the generalized calculation taking into account tilts of the optics and transverse misalignment \cite{Tovar1995} could also not explain this features, which suggests that other effects should be taken into account, such as  wave-front distortions introduced by the optical elements. Such a study requires  a more advanced modeling   of our resonator, which will be the subject of the next section \ref{sec:aberrations}.

\subsubsection{Variation of the optical gain with $\delta_2$}
\label{subsec:variation_gain_d2}

\begin{figure}[!h]
\centering
\includegraphics[scale=0.4]{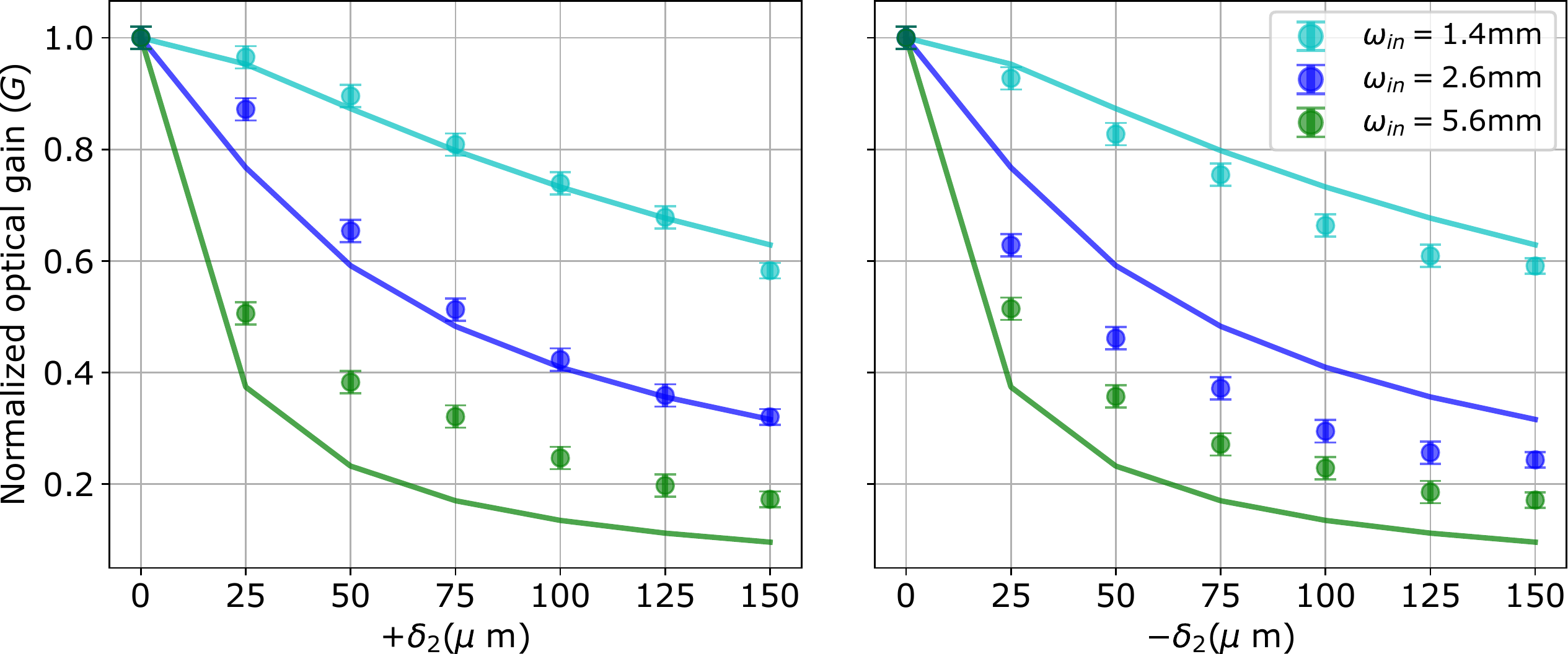}
\caption{Relative variation of the optical gain  with respect to  $\delta_2$ for  various input beam sizes $w_{in}$. 
 Circular markers:  experimental data. Solid lines: calculation using the ABCD transfer matrix method. 
  Left:  $\delta_2>0$  (increase in cavity length). Right:  $\delta_2<0$  (decrease in cavity length).}
\label{fig:gain_delta_peak}
\end{figure}

 Fig.~\ref{fig:gain_delta_peak} shows the relative variation of the optical gain  with $\delta_2$. The relative optical gain is defined as the maximum of the resonance curve (as shown in Fig.~\ref{fig:resonances}) for a given $\delta_2$ relative to the maximum of the resonance for $\delta_2=0$.
For each value of $\delta_2$, six resonance curves  were recorded and the maximum of the peaks was extracted; the mean and standard deviation were then obtained from these six measurements. 
 The decrease in optical gain is well captured by the ABCD matrix calculation for the small input beam size $w_{in}=1.4$~mm (cyan plain lines), but fails to quantitatively reproduce the experimental data at larger beam sizes ($w_{in}=$2.6~mm and 5.6~mm). 
This mismatch further supports a more advanced modeling of our cavity, as will be presented in the next section.

\subsection{Influence of optical aberrations on the optical gain and beam shape}
\label{sec:aberrations}

\subsubsection{Maximum optical gain and finesse}
\label{subsec:gain_finesse}
\begin{figure}[!h]
\centering
\includegraphics[scale=0.45]{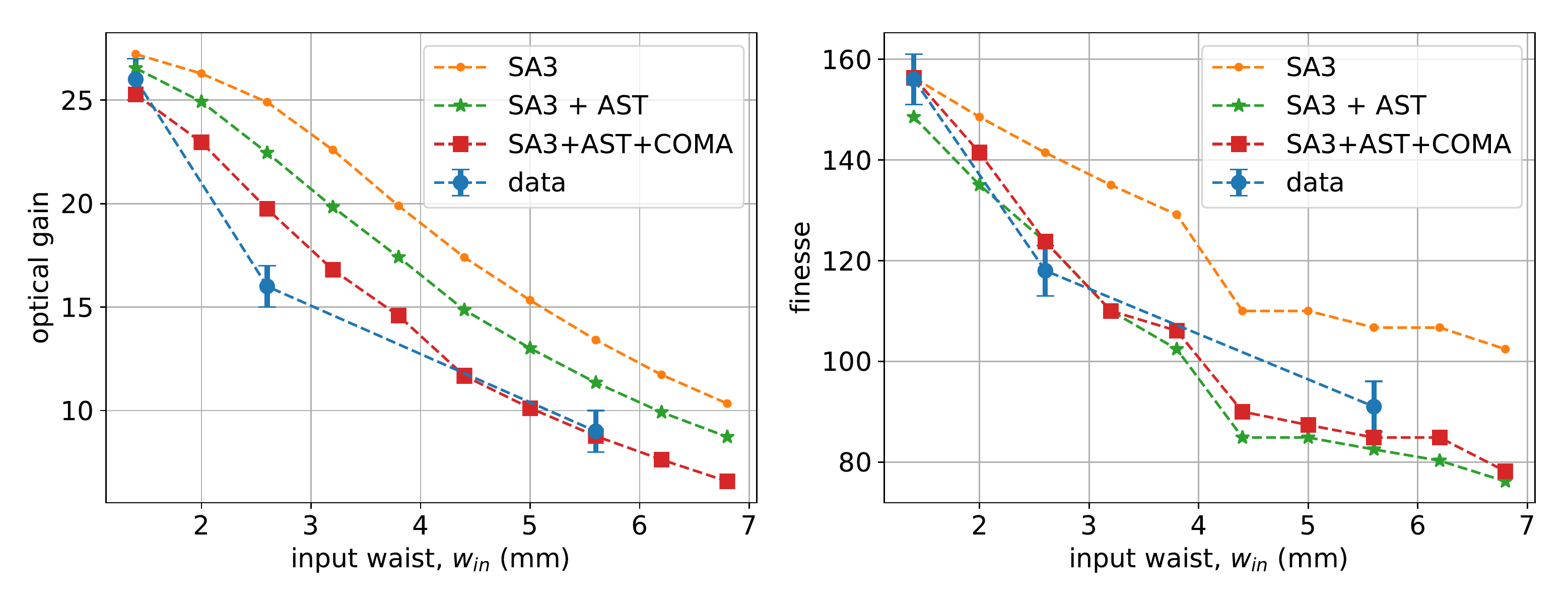}
\caption{Dependence of the optical gain and the finesse of the resonator on the waist of the input beam, $w_{in}$. Blue filled circles: experimental data (mean and standard deviation obtained from these six measurements). The other markers show the result of numerical calculations using the ASM technique, tacking into account different optical aberrations  of the optical elements, as measured in table \ref{table_Zcoeff}. Orange dots: third order spherical aberration (SA3); green stars: SA3 and astigmatism (AST); red squares: all measured aberrations.}
\label{fig:gain}
\end{figure}

We now move to the study of the finesse and of the absolute optical gain   of the resonator when it is aligned in its best configuration, i.e. at the closest point to the ideal $f-f$ case that is experimentally achievable (i.e. $\delta_2\simeq0$).
From the power measurements of $P_{in}$ ($P_{out}$)  at the input (output) of the resonator, we estimate the  optical gain  as $G={(P_{out}/T_2)}/{P_{in}}$, where $T_2=1-r_2^2$ is the transmission coefficient of M2. 
The output power  was measured by slowly varying the cavity length across resonance and tracking the values with a power meter. The measurement was repeated three times for each value of input beam waist, $w_{in}$
The left panel of Fig.~\ref{fig:gain} shows the variation of the optical gain with respect to the input beam waist, $w_{in}$, where it can be observed that the measured gain (blue filled circle) decreases while increasing $w_{in}$. The finesse follows a similar trend, as shown on the right panel of Fig.~\ref{fig:gain}.
These observations are not supported by the  ABCD transfer matrix calculation, which predicts a constant value for the gain and finesse (Eq.~\eqref{eqn:GF}) for all values of $w_{in}$.

\begin{table}[!h]
\begin{center}
\begin{tabular}{|c|c|c|c|c|c|}
\hline 
Element & \multicolumn{2}{|c|}{ Astigmatism } & \multicolumn{2}{|c|}{ Coma } & Spherical aberration \\
 & \multicolumn{2}{|c|}{ ($\pm0.001$) } & \multicolumn{2}{|c|}{($\pm0.02$)  } &   3$^{rd}$ order ($\pm0.05$)\\ 
\hline 
Lens & 0.049 & 0.02 & 0.155 & 0.138 & 1.26 \\ 
\hline 
M2& 0.015 & 0.029 & 0.012 & 0.002 & 0.01 \\ 
\hline 
M1 & 0.010 & 0.007 & 0.006 & 0.032 & 0.06 \\ 
\hline 
\end{tabular}
\end{center} 
\caption{Characterization of the wavefront distortion caused by the   optical elements of the resonator. We show here the coefficients of the Zernike decomposition of the wavefront, as analyzed with a ZYGO  interferometer on the full clear aperture of the optics (45~mm). The coefficients are given in units of $\lambda=852$~nm. 
The two values for Astigmatism and Coma give the two axes used in the decomposition of the wavefront,  corresponding to the Zernike orders (2,2) [Astigmatism at 0 degree], (2,-2) [Astigmatism at 45 degrees], (3,1) [Coma at 0 degree] and (3,-1) [Coma at 45 degrees]. The third order spherical aberration correspond to the Zernike order (4,0).}
\label{table_Zcoeff} 
\end{table}

The failure of the ABCD matrix propagation to reproduce our data  called for an improved model of our resonator taking into account the imperfections of the optical elements.
To that purpose, we developed a set of numerical calculations based on the angular spectrum propagation of fields, with  the same numerical methods  as in the OSCAR cavity simulation software \cite{Degallaix2010}. The details of the numerical calculations are given in appendix~\ref{sec:appendixCalculations}.
As described in Ref.~\cite{Kozacki2008a}, the angular spectrum method (ASM) is  limited by computational memory constraints derived from the grid size and the number of round trips to consider (about $2\times \mathcal{F}\simeq 350$ round-trips for our resonator). In our application,  simulating beams with larger input waists  translates into the need to use  a  calculation grid with a smaller step in order to sample correctly the smaller beam sizes at $M_2$. 
At first, to overcome this issue in the particular case of problems with cylindrical symmetry, we performed ASM calculations  using the Hankel transform \cite{Pritchett2001}.

\textbf{Effect of spherical aberration.}
We measured the optical aberrations of each optical component of our resonator using a ZYGO interferometer, see Table \ref{table_Zcoeff}. 
As expected, the third order spherical aberration is the greatest source of wavefront distortion. 
Studying the influence of optical aberrations on the buildup of the resonance of a beam in the cavity is computationally less demanding in the cylindrically symmetric case, where the Hankel transform can be exploited.

The orange dots in Fig.~\ref{fig:gain} present the results of the numerical calculation based on the ASM taking into account the measured third order spherical aberration (SA3) of the lens ($1.3 \lambda$ over 45~mm). Numerically, the distance $\delta_2$ was optimized for each value of the waist to find the resonance with the maximum optical gain (see appendix~\ref{sec:appendixCalculations} for details). 
Fig.~\ref{fig:gain} shows that  the reduction in gain and finesse is qualitatively well reproduced  by the numerical calculation taking into account SA3 only, but not quantitatively. 
Not shown in the figure, we also numerically varied the value of the aberration coefficient (around $1.3\lambda$) to estimate the impact on the gain and finesse: for example, at the waist value $w_{in}=6.8$~mm, a variation of $8\%$ of the aberration coefficient results in a variation of the optical gain of $2\%$.
For small beam waists compared to the clear aperture radius of the lens, we also found that the value of $\delta_2$ yielding the maximum gain  evolves linearly with the value of the SA3 coefficient. This can be explained by the fact that the SA3 is mostly a pure curvature in the center of the optics (i.e. an effective focal length change), which can be compensated by adjusting $\delta_2$.

\textbf{Effect of other aberrations.}
In order to calculate the impact of aberrations without cylindrical symmetry, such as astigmatism or coma as reported in table \ref{table_Zcoeff}, a full  2D calculation based on the ASM  is required (in contrast to using a quasi-1D method with the Hankel transform as above).
The basic implementation of the ASM calculation introduces severe computer memory constraints: for example, a correct sampling of the optical field for an input waist of 5.6~mm requires a  calculation   grid with at least $2^{11}\times 2^{11}$ points, which, for the propagation of 300 round trips, occupies 10 GBytes of memory. 

To relax the constraints on computer memory, we  implemented the so-called method of \textit{propagation with magnification} \cite{sziklas_mode_1975,vinet_virgo_2006}, which tackles the propagation of converging or diverging beams of light by transforming them into equivalent quasi-collimated optical beams in a new coordinate system. This method can only be applied when the paraxial approximation is considered valid, which is fulfilled for most of the waist values considered in this work: for $w_{in}=5.6$~mm, the Fresnel number $w_{in}^2/\lambda f\simeq 167$ is at the limit of this approximation. Nevertheless, we have verified that the regular ASM and the implementation of the propagation with magnification yield the same results here. The details about the method and its implementation are given in appendix~\ref{subsec:prop_with_magnif}. The implementation of this method allows us to reduce by  16 the number of points to use for a proper calculation, i.e. a reduction of 16 on the necessary computer memory; for example, the propagation of a beam with input waist of 5.6~mm can be performed with $2^9\times 2^9$ points. The use of the method of propagation with magnification allowed us to better cover the parameter space (e.g. varying the aberrations and the beam waist) and to obtain quantitative results by propagating the beam over 350 round trips in the cavity.
In order to go beyond the paraxial approximation imposed by the method of propagation with magnification (i.e. for large $w_{in}$ in this work), alternative ASM simulation techniques could be adopted to generate arbitrary sampling in the diffraction plane, as mentioned in Refs.~\cite{Xiao2012,Jurling2018}, still with the same goal to save computing resources.


The results of the calculations taking into account the different types of aberrations are given in Fig.~\ref{fig:gain}: the green stars (red squares) show the effect of astigmatism (astigmatism and coma) on the gain and finesse, additionally to the SA3. Remarkably, the calculation tacking into account all the measured aberrations (red squares) reproduce quantitatively the experimental data without free parameters. This confirms the critical impact of the aberrations on the performance of the resonator in the degenerate case. 
Such quantitative agreement will allow us to design the resonator in future studies, i.e.  to put requirements on the wave-front distortions introduced by the optics to reach given performances.

\begin{figure}[!h]
\centering
\includegraphics[width=\linewidth]{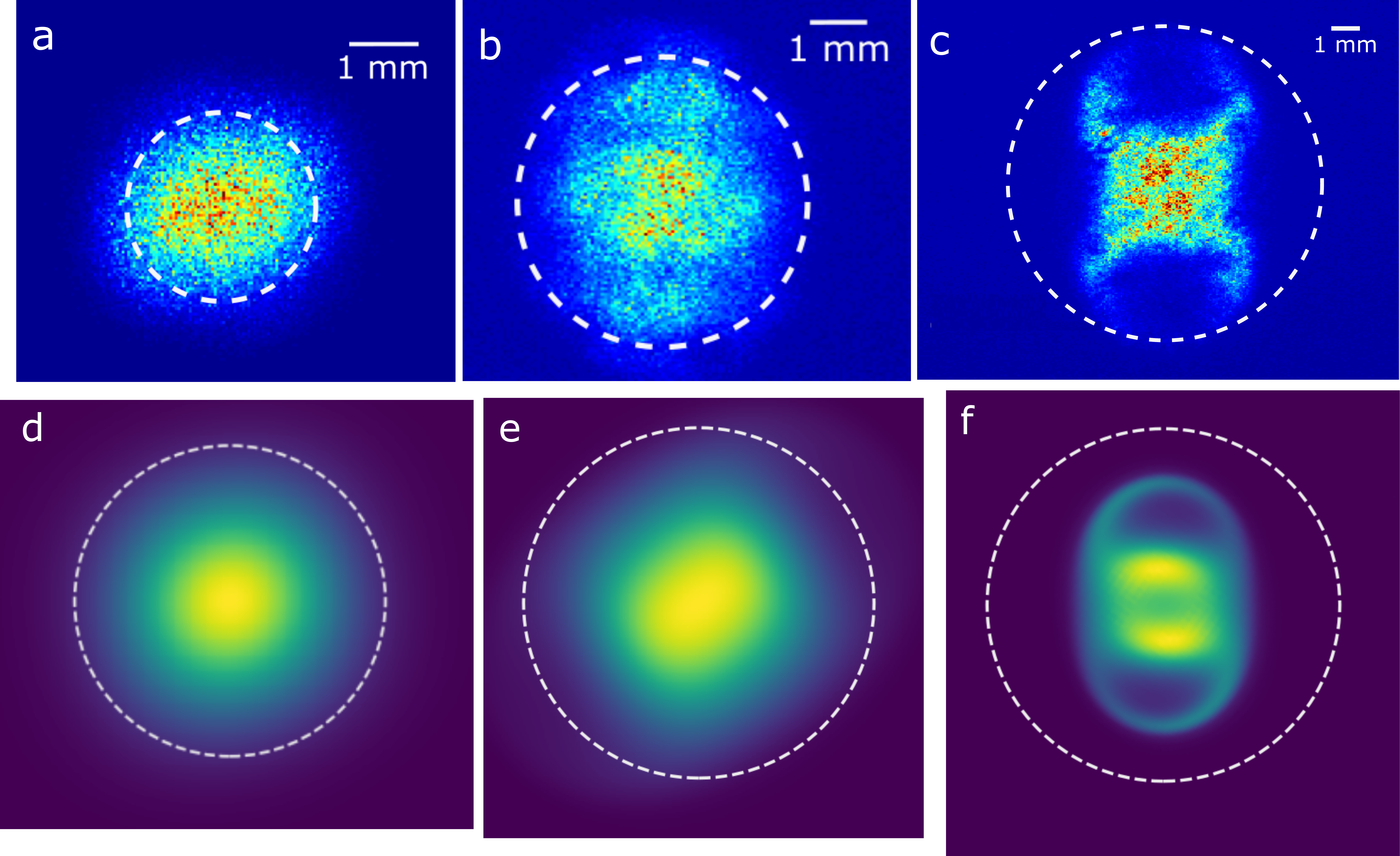}
\caption{ Pictures of the transmitted beam for $w_{in}=1.4$~mm (a), $w_{in}=2.6$~mm (b) and $w_{in}=5.6$~mm (c). 
(d), (e) and (f) show the corresponding calculated profiles on resonance, using the angular spectrum method and taking into account the aberrations measured in Table~\ref{table_Zcoeff}. In all the panels, the dashed circles indicate the $1/e^2$ diameter of the input beam. }
\label{fig:beam_shape}
\end{figure}

\subsubsection{Spatial profile of the resonating beam}
\label{subsec:shape}
We recorded the profile of the resonating beams using a  CCD camera in transmission of the cavity (see Fig.~\ref{fig:schem}). The CCD camera images a paper screen onto which falls the beam transmitted from $M_2$ after a known distance, with a calibrated magnification.
 To observe the beam shape along the resonance, we slowly scanned the resonance profile with a typical period of 10~seconds per free spectral range and saved several pictures.
Figs.~\ref{fig:beam_shape}(a), (b) and (c) show the profile of the  beam at resonance, respectively for  $w_{in}$ = $1.4$~mm, $2.6$~mm and $5.6$~mm. 
While the beam shape is Gaussian for $w_{in}$ = $1.4$~mm (a), structural deviations with respect to the shape of a Gaussian profile gradually appear when increasing the input waist size, and they become severe for  $w_{in}=5.6$~mm (c). 
These structural deviations are qualitatively reproduced by the numerical calculations (panels (d,e,f)) using the aberrations measured in table \ref{table_Zcoeff}.

\section{Resonance of a top-hat beam}
In the perfectly aligned cavity, i.e. when the resonator is rigorously degenerate, a  beam with an arbitrary intensity profile can  resonate, in principle. This is true, in particular, for a beam with a flat intensity profile in a given region, the so-called top-hat or flattop beam \cite{Gori1994}. Nevertheless, the effects of misalignment or aberrations of the optics will affect the resonance of such a beam,  with an impact possibly more severe than for a Gaussian beam of similar diameter. Indeed, a top-hat beam can be decomposed over a basis of several Gaussian beams \cite{Gori1994}, whose resonance will be impaired by such imperfections, as studied in the previous sections. Here, we use our numerical calculation tools to theoretically study the propagation of a top-hat beam in the presence of misalignment or aberrations, having in mind the application of a top-hat optical resonator for atom interferometry \cite{Mielec2018}.

\begin{figure}[!h]
\centering
\includegraphics[width=\linewidth]{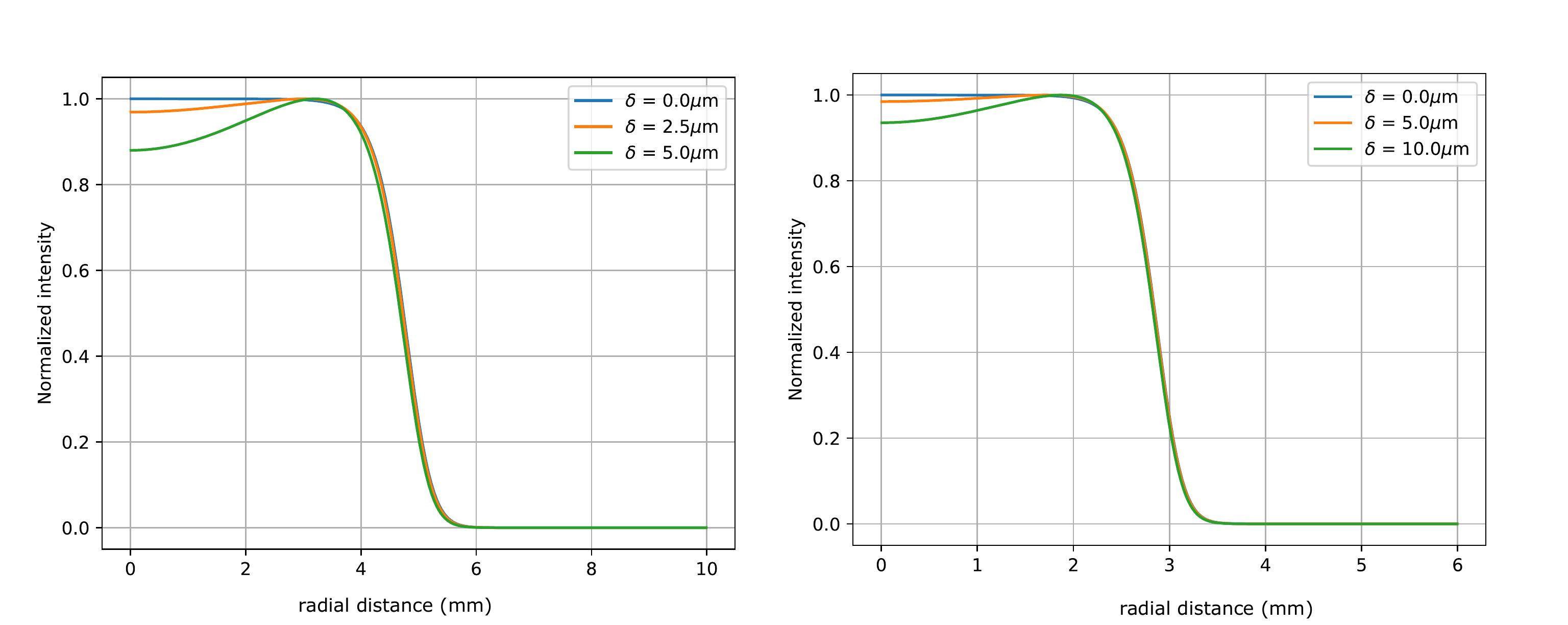}
\caption{Effect of longitudinal misalignment $\delta_2$ on the propagation of a top-hat beam with $r_0=5$~mm (left) and $r_0=3$~mm (right).}
\label{fig:top_hat_d2}
\end{figure}

We first concentrate on the influence of longitudinal misalignment on the build-up of the top-hat beam characterized by a cylindrically symmetric Fermi-Dirac intensity distribution. Fig.~\ref{fig:top_hat_d2} shows the influence of varying $\delta_2$ on the profile of the resonating beam, obtained from numerical calculation using the Hankel transform, for top-hat beams  of two different characteristic  radii: $r_0=5$~mm (left) and $r_0=3$~mm (right). Since the top-hat beam of larger radius decomposes on a basis of Gaussian beams of larger waist, a stronger deformation is observed for a given value of $\delta_2$, consistent with the results found in the previous section.
A control of longitudinal misalignment at this level will be necessary to ensure the proper propagation. 
 Regarding the effect on the optical gain, the range of $\delta_2$ values investigated here has little influence: for $r_0=3$~mm and $\delta_2=5 \ \mu$m, the gain is still 27.3, close to the theoretical limit imposed by the finite transmission and reflection coefficients of the lens and  mirrors, respectively (Eq.~\eqref{eqn:GF}).

\begin{figure}[!h]
\centering
\includegraphics[width=\linewidth]{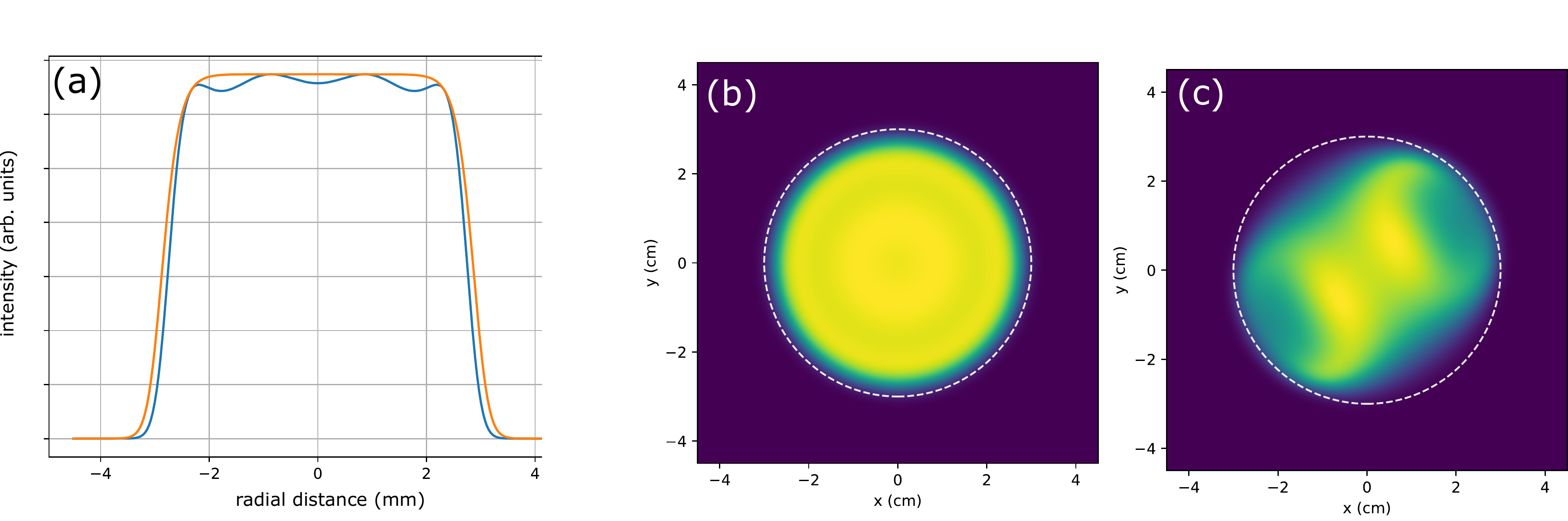}
\caption{Effect of optical aberrations on the propagation of a top-hat beam with $r_0=3$~mm. Panels (a) and (b) show the profiles taking into account third order spherical aberration only, while panel (c) additionally incorporates coma and astigmatism. Panel (a) is a line cut of panel (b).}
\label{fig:top_hat_aberrations}
\end{figure}

We then show in Fig.~\ref{fig:top_hat_aberrations} the effect of aberrations of the optics on the spatial profile and on the gain of a top-hat beam with $r_0=3$~mm. First, we present separately the influence of third order spherical aberrations as investigated in the previous section, in panels (a) and (b) displaying  a radial cut and the full two-dimensional radial cross-section, respectively. Here, the distance of the mirror M2 has been matched to find the best resonance, i.e. to compensate the pure curvature effect linked to the central part of the spherical aberration. We numerically find a gain of 25.1 and a finesse of 135. Panel (c) shows the effect of the other aberrations, namely astigmatism and coma, with the same parameters as in table \ref{table_Zcoeff}. For this calculation, we used the propagation with magnification method (see previous section). The gain is 18.1 and finesse 114. 

This study shows that the effect of the third order spherical aberration does not significantly alter the shape of the beam, in opposition to the  non cylindrically-symmetric wavefront distortions, which shall be considered carefully in the design.

\section{Discussion and outlook}

We have studied a 44-cm long, linear, degenerate optical resonator consisting of two flat mirrors and a lens in a regime of moderate finesse ($\sim 150$), in which laser beams with waists of varying size up to 5.6~mm  resonate. We described the alignment procedure, which requires particular care due to the degeneracy of the cavity. The influence of longitudinal misalignment of tens of $\mu$m has been investigated and, in the regime of intermediate beam sizes (waists smaller than 2~mm),  is well captured by a model based on ABCD matrix propagation of Gaussian beams.

Deviations from the behavior expected by the ABCD model for waists larger than 2~mm called for a study of the impact of optical aberrations. Our model, based on the angular spectrum method and tacking into account the measured wave-front distortions of our optical elements,  reproduces quantitatively the decrease of the optical gain when  beams with increasing  waists are injected in the cavity, and  reproduces qualitatively the observed distortions in the beam shape. To model  our resonator, we developed different numerical tools which tackle the computer memory constraints associated with the large variation of beam size during propagation in the cavity. Such numerical tools are useful to study the impact of aberrations or misalignment on the performance of the cavity and provide guidance in the alignment of the resonator. The  tools and  methods presented in this work are applicable to different cavity geometries. They can, for example, be of interest  for the simulation of optical propagation in gravitational wave detectors, in complement of existing methods such as \textit{Finesse} \cite{gwoptics}.


Our work was motivated by the potential interest of using  an enhancement cavity for building compact cold-atom interferometers, where laser beam diameters of several millimeters  are favorable. In that context, reaching an optical gain of 26 for a waist of 1.4~mm already constitutes a useful result. Our cavity could for example be used to drive an atom interferometer with large momentum transfer beam splitters, which require high optical powers \cite{Mueller2008,Leveque2009}. 
Optical enhancement cavities for atom interferometry are also interesting in the context of future low-frequency gravitational wave detectors \cite{Canuel2018}.
Moreover, owing to the degeneracy of our resonator, beams of arbitrary spatial shape can  resonate, in principle. In that context, we numerically studied the effect of misalignment and aberrations of the optics on the resonance of a beam with a top-hat intensity profile,  which has been shown to be beneficial for cold-atom interferometry applications \cite{Mielec2018}. 


\section*{Acknowledgments.}
\label{Acknowledgments}
We thank Walid Chaibi, Yvan Sortais, Nir Davidson and Andreas Freise for stimulating discussions. We acknowledge Institut d'Optique (Palaiseau) for lending us the ZYGO interferometer.

\section*{Funding.}
This work was supported by the European Union's Horizon 2020 research and innovation programme under the Marie Sklodowska-Curie grant agreement No 660081, by Ville de Paris (project HSENS-MWGRAV), Centre National d'Etudes Saptiales (CNES), FIRST-TF (ANR-10-LABX-48-01), Sorbonne Universit\'es (project LORINVACC) and ANR (project PIMAI, ANR-18-CE47-0002-01). 

\section*{Disclosures.}
The authors declare no conflicts of interest.

\section{Appendix: details of the numerical calculations}
\label{sec:appendixCalculations}

\textbf{Angular Spectrum Method.}
The Python codes used for the numerical calculations used in this paper are available on GitHub at this reference \cite{remigeiger_2020_4011610}. We give here the most important concepts of the calculations.

To compute the total field inside the cavity, we propagate the input field using physical optics (Angular Spectrum Method, ASM, Ref.~\cite{Degallaix2010}) for a certain number of round trips.
The number of round trips is determined by the finesse $\mathcal{F}$ of the cavity; we typically make the calculations for  $2\mathcal{F}\simeq 350$ round trips.
The total field is then the complex sum of the propagated fields corresponding to each round-trip.


Once the fields are calculated, we look for the condition of constructive interference by numerically scanning the resonance.
To this aim, we introduce  a given phase shift $\phi$ per round-trip, which accumulates with round-trips ($\phi$ for first round-trip, $2\phi$ for second, and so on).
This introduced phase shift simulates a change of frequency of the input laser frequency (on the order of magnitude of the free spectral range), or a variation of the length of the resonator as performed in the experiment with the PZT on mirror M2 (length variations of the order of half of the wavelength).
Since this introduced phase shift is small (i.e. associated to length variations $\delta_2\sim \lambda/2\simeq 400 \ \mathrm{nm}\ll 10 \ \mu\mathrm{m}$, to be compared with the scale of Fig.~\ref{fig:gain_delta_peak}), we assume  that the spatial shapes of the field are not modified by the introduction of such a phase shift, which we numerically verified.
This assumption allows us to compute the propagated fields in a first step, and then to scan the resonance of the cavity in a second step, in order to pinpoint the position of the maximum optical gain.

\textbf{Propagation with magnification.}
\label{subsec:prop_with_magnif}
When attempting to propagate light fields using the angular spectrum method, one can encounter a number of numerical problems (e.g. aliasing). A fine space-discretization is required to sample the $(x, y)$ space transverse to the propagation direction. 
One possible way to overcome these problems, or at least lower their effect, is to use the propagation algorithm presented by Sziklas et al. \cite{sziklas_mode_1975}, called the propagation with magnification method. This method tackles the propagation of converging or diverging beams of light by transforming them into equivalent quasi-collimated optical beams in a new coordinate system based on Gaussian beam theory. This method can only be applied when the paraxial approximation is  valid. We briefly present here the method and its implementation.

Let $E(x, y, z)$ be a field supposed to be a solution of the paraxial equation
\begin{equation*}
    \dfrac{\partial ^2 E}{\partial x^2} + \dfrac{\partial ^2 E}{\partial y^2} - 2ik\dfrac{\partial E}{\partial z} = 0,
\end{equation*}
with $k = 2 \pi / \lambda$ the wavevector of the light. The transformation
\begin{equation}
    F(x, y, z) = z \exp\left(i \dfrac{k (x^2 + y^2)}{2z}\right) E(x, y, z)
    \label{Fdef}
\end{equation}
is then performed and it is shown in Ref.~\cite{sziklas_mode_1975} that the introduced function $F$ is a solution of the paraxial equation in a new system of coordinates $\{x', y', z'\}$ . These new coordinates are defined by 
\begin{equation}
\begin{cases}
    &x' = \dfrac{\alpha x}{z}, \\
    & \\
    &y' = \dfrac{\alpha y}{z}, \\
    & \\
    &z' = \dfrac{\alpha^2 (z-z_0)}{zz_0},
    \label{coordinates}
\end{cases}
\end{equation}
with $\alpha$ and $z_0$ being positive constants as chosen below. Hence, we have
\begin{equation}
    \dfrac{\partial ^2 F}{\partial x'^2} + \dfrac{\partial ^2 F}{\partial y'^2} - 2ik\dfrac{\partial F}{\partial z'} = 0.
    \label{sziklas}
\end{equation}

\begin{figure}
    \centering
    \includegraphics[scale = 0.4]{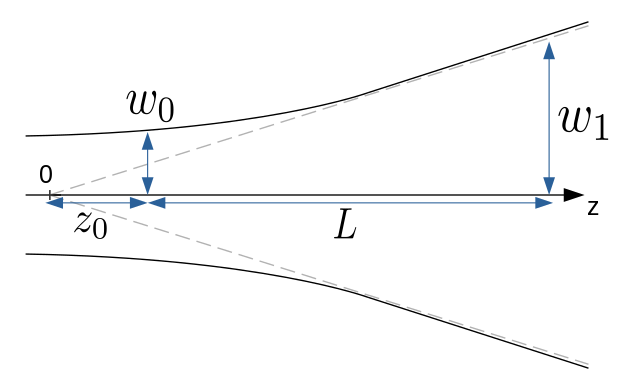}
    \caption{Sketch of the propagating beam and illustration of the constants used in the propagation method.}
    \label{fig:diverging}
\end{figure}

From Eq.~\eqref{sziklas}, we can introduce the propagation with magnification algorithm. We consider a diverging beam like the one sketched on Figure \ref{fig:diverging} (the converging beam case is treated in a same way). The field propagates from a point $z = z_0$ over a distance $L$, $w_0$ is the beam waist, and $w_1$ is the waist after the propagation over the distance $L$. As pointed out by J.Y. Vinet in Ref.~\cite{vinet_virgo_2006}, setting the constants in Eq.~\eqref{coordinates} to 
\begin{equation*}
    z_0 = \dfrac{L}{\left(w_1/w_0 - 1\right)}, \quad \alpha = z_0,
\end{equation*} 
allows the primed coordinates to expand at the same geometrical ratio as the diverging beam. In that way, the function $F$ defined by Eq.~\eqref{Fdef} represents a quasi-collimated beam in the $(x',y')$-space.

From now on we assume the $(x,y)$ space to be discretized in $N\times N$ cells. We suppose that we know the field $E(x, y, z_0) = E_0(x,y)$ before propagation.

The algorithm goes through the following steps:
\begin{enumerate}
    \item We define the function $F_0$ as
    \begin{equation*}
        F_0(x,y) = z_0 \exp{\left(i\dfrac{k(x^2 + y^2)}{2z_0}\right)} E_0(x,y).
    \end{equation*}
    
    \item We propagate the resulting beam using the usual propagator of the ASM in the paraxial approximation, but this time in the $\{x', y', z'\}$ coordinates, because $F_0$ is solution of the paraxial equation in that coordinate system. As in the standard ASM propagation method, we apply the propagator in the Fourier space, which consists in the transformation
    \begin{equation*}
        F_1(x',y') = \mathcal{F}^{-1}\left[\mathcal{F}\left[F_0(x,y)\right] \exp\left( i \dfrac{2\pi^2 (\nu_{x'}^2 + \nu_{y'}^2) \Delta z'}{k} - ikL \right)\right],
    \end{equation*}
    with $\Delta z' = w_0 L / w_1$, $\nu_{x',y'}$ the spacial frequency in primed coordinates, and $\mathcal{F}[\cdot]$ denoting the Fourier transform.
    \item We cancel the action on the phase of the first step by defining $E_1$ as the propagated field in the primed coordinates
    \begin{equation*}
        E_1(x',y') = \dfrac{1}{z_0+L} F_1(x',y') \exp\left(- i \dfrac{k(x'^2 + y'^2)(z_0 + L)}{2 z_0^2}\right).
    \end{equation*}
    \item The last step is to go back to the normal transverse coordinate system:
    \begin{equation*}
        x = \dfrac{w_1}{w_0}x',\quad y = \dfrac{w_1}{w_0}y'.
    \end{equation*}
\end{enumerate}

A guess value for $w_1$ shall be given  for the algorithm to work.  It has been numerically observed that overestimating the waist by two or three times has no influence on the result. 
Nonetheless, underestimating $w_1$ has dramatic effects: indeed, when using a value of $w_1$ smaller than the expected size, the beam size increases and soon goes beyond the calculation grid, which leads to diffraction. This situation should therefore be avoided.

We verified that the regular ASM implementation and that the propagation with magnification give the same results in several cases, and coincide with the results of the method using the Hankel transform for wave-front distortions with cylindrical symmetry.

\label{biblio}

\end{document}